\documentclass[twoside,english,journal]{IEEEtran}
\usepackage[latin9]{inputenc}
\usepackage{verbatim}
\usepackage{graphicx}

\makeatletter
\usepackage{booktabs}
\usepackage{graphicx}

\title{Intermodular Configuration Scrubbing of On-detector FPGAs for the ARICH at Belle II}
\author{R.~Giordano, Y.~Lai, S.~Korpar, R.~Pestotnik, A.~Lozar, L.~\v{S}antelj, M.~Shoji, and S.~Nishida
\thanks{Corresponding author: R. Giordano (email: raffaele.giordano@unina.it)}
\thanks{R.~Giordano is with Università degli Studi di Napoli "Federico II" and INFN, 80126, Italy}
\thanks{Y.~Lai, M.~Shoji and S.~Nishida are with High Energy Accelerator Research Organization (KEK), 305-0801, Japan}
\thanks{S.~Korpar is  with University of Maribor, 2000, Slovenia and Jo\v{z}ef Stefan Institute, 1000, Slovenia}
\thanks{R.~Pestotnik and A. Lozar are with Jo\v{z}ef Stefan Institute, 1000, Slovenia}
\thanks{L.~\v{S}antelj is with University of Ljubljana and Jo\v{z}ef Stefan Institute, 1000, Slovenia}
}

\makeatother

\usepackage{babel}
\begin{document}
\maketitle 
 \thispagestyle{empty} \pagestyle{empty} \renewcommand{\figurename}{Fig.} \renewcommand{\tablename}{TABLE}
\begin{abstract}
On-detector digital electronics in High-Energy Physics experiments
is increasingly being implemented by means of SRAM-based FPGA, due
to their capabilities of reconfiguration, real-time processing and
multi-gigabit data transfer. Radiation-induced single event upsets
in the configuration hinder the correct operation, since they may
alter the programmed routing paths and logic functions. In most trigger
and data acquisition systems, data from several front-end modules
are concentrated into a single board, which then transmits data to
back-end electronics for acquisition and triggering. Since the front-end
modules are identical, they host identical FPGAs, which are programmed
with the same bitstream. 

In this work, we present a novel scrubber capable of correcting radiation-induced
soft-errors in the configuration of SRAM-based FPGAs by majority voting
across different modules. We show an application of this system to
the read-out electronics of the Aerogel Ring Imaging CHerenkov (ARICH)
subdetector of the Belle2 experiment at SuperKEKB of the KEK laboratory
(Tsukuba, Japan). We discuss the architecture of the system and its
implementation in a Virtex-5 LX50T FPGA, in the concentrator board,
for correcting the configuration of up to six Spartan-6 LX45 FPGAs,
on pertaining front-end modules. We discuss results from fault-injection
and neutron irradiation tests at the TRIGA reactor of the Jožef Stefan
Institute (Ljubljana, Slovenia) and we compare the performance of
our solution to the Xilinx Soft Error Mitigation controller. 
\end{abstract}


\section{Introduction}

\IEEEPARstart{O}{n-detector} digital electronics in High-Energy Physics
(HEP) experiments is increasingly being implemented by means of Static
Random Access Memory-based (SRAM-based) Field Programmable Gate Arrays
(FPGAs) \cite{Xilinx_Ultrascaleplus,Intel_Agilex}. The main reasons
are that these devices are reconfigurable, they are capable to process
large amounts of data in real-time and to perform multi-gigabit data
transfers on serial lines. Radiation-induced single event upsets (SEUs)
in the device configuration hinder the correct operation, since they
may alter the programmed routing paths and logic functions \cite{Wirthlin_HiRel,QuinnRadEffectsFPGAs}.
{} These errors need to be removed, i.e. scrubbed \cite{HybridScrubbing},
as soon as possible. If accumulated, they can even break triple modular
redundancy (TMR) schemes \cite{TMR_Accumulation2}. Simple scrubbing
schemes foresee additional radiation-hardened memories for storing
a golden bitstream, so they make it possible to correct any number
of upsets per configuration frame (i.e. the smallest accessible configuration
element). Other solutions exploit error correcting codes, such as
the Xilinx Soft Error Mitigation (SEM) controller, and they make it
possible to correct few upsets per frame. 

Recently, novel scrubbing techniques based on redundancy of configuration
frames have been proposed \cite{Frame_Level_Redundancy,Configuration_SelfRepair}
and they make it possible at the same time to avoid external memories
and have no \textit{a priori} limit on the number of correctable errors.
These techniques require to generate redundant configuration frames
in the device and to provide circuits to majority vote frames for
data detection and correction.%

In most trigger and data acquisition systems, data from several front-end
modules are concentrated into a single board, which then transmits
data to back-end electronics for acquisition and triggering, as for
instance in \cite{ATLAS_Muon,Belle2_TOP}. The front-end modules are
identical and their FPGAs are programmed with the same bitstream,
which can be uploaded via the data concentrator board. 

The contribution of this work to the state of the art is twofold.
On one hand, we present a novel scrubber which majority votes configuration
frames of FPGAs across different modules. The main advantage of our
solution is that there is no impact on the resource occupation in
the device for generating the redundant frames, since the inherent
redundancy of different modules is leveraged. 
\begin{figure}
\begin{centering}
\includegraphics[width=0.9\columnwidth]{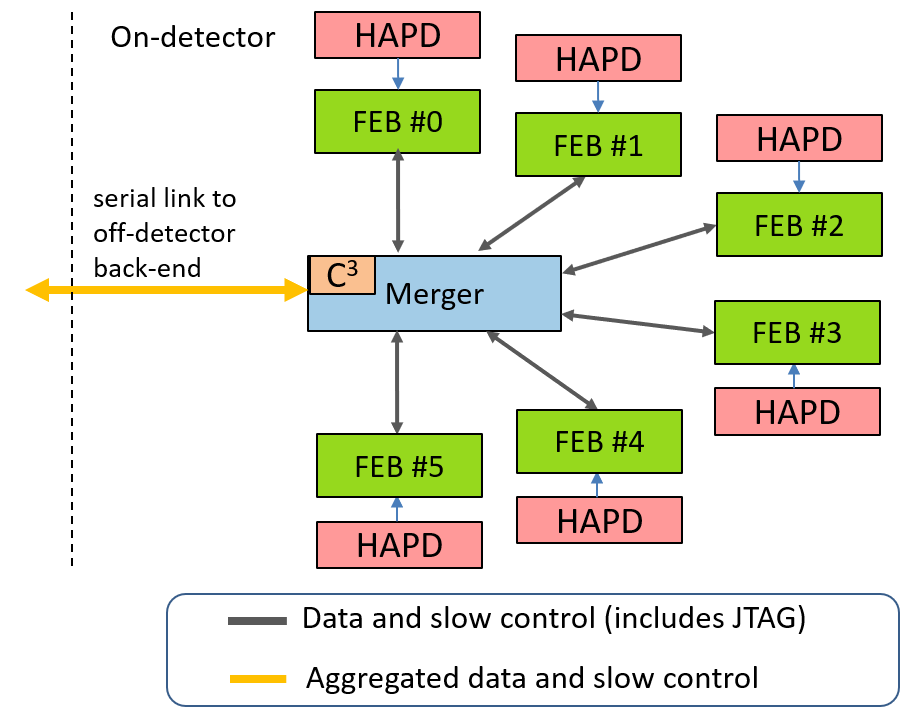}
\par\end{centering}
\caption{\label{fig:ARICH_blockDiagram}Simplified diagram of one merger and
six front-end boards set in the ARICH readout electronics.}
\end{figure}
 On the other hand, we show an actual case study of our concept in
a running experiment. In fact, we applied our concept to the Aerogel
Ring Imaging CHerenkov (ARICH) counter of the Belle II \cite{Belle2_exp}
experiment at the SuperKEKB $e^{+}e^{-}$collider (Tsukuba, Japan). 

\section{The ARICH counter}

The ARICH is part of the crucial particle identification (PID) system
\cite{importance_of_PID}, which is required for B-meson flavor tagging
in CP violation studies in the neutral B system. PID is also key in
precision measurements of rare B and D decays, since it makes it possible
to suppress backgrounds. The ARICH consists of 124 pairs of aerogel
tiles as radiators, an array of 420 144-channel Hybrid Avalanche Photo-Detectors
(HAPDs), and the pertaining readout system \cite{ARICH_frontend}.
A front-end board (FEB) is attached to each HAPD and it hosts four
application specific integrated circuits (ASICs) and a Spartan-6 LX45
FPGA. Groups of six (or in a few cases five) FEBs transfer digitized
hit information to a merger board, built around a Virtex-5 LX50T FPGA,
which transmits data to the off-detector electronics (Fig. \ref{fig:ARICH_blockDiagram}).
Spartan-6 devices are produced with a high concentration of Boron
as a p-type dopant. The high cross-section of $^{10}B$ for thermal
neutron capture leads to an increased SEU rate with respect to other
Xilinx devices. Irradiation tests at the TRIGA reactor \cite{TRIGA}
of the Jožef Stefan Institute (Ljubljana, Slovenia) made it possible
to measure the configuration upset rate with a neutron spectrum similar
to one of the Belle II spectrometer. The results extrapolated to Belle
II conditions provided a rate of 8 SEUs per hour per board, or 3.3k
SEUs per hour overall. %
{} Single-bit errors per frame can be recovered by the SEM, but multi-bit
errors are unrecoverable. Our results have also shown that the SEM
cannot effectively limit the accumulation of SEUs in the Spartan-6
configuration, therefore, we decided to address this issue by designing
a custom solution.

\section{The Configuration Consistency Corrector}

The scrubber we designed, named Configuration Consistency Corrector
($C^{3}$), operates in the merger-board FPGA for majority voting
the configuration of up to 6 FEB FPGAs. It is built around a Xilinx
picoBlaze 6 (pB) soft-core, it runs at 127 MHz (frequency used by
the Belle II link system) and it features parallel readback for the
target FPGA configuration. The JTAG IO can be performed in two modes:
single and broadcast read/write. Single mode makes it possible to
write to and read from a single FPGA, while broadcast mode permits
simultaneous write to and read from multiple FPGAs, in a majority-voted
fashion. Block RAMs (BRAMs) are used for storing the pB program, configuration
frames read from or to be written to target FPGAs, and device-specific
details about the frame address increment logic. Finally a UART (or
optionally a JTAG interface) makes it possible to send commands to
the core and to log details about the detected SEUs (device, frame
address, bit offsets, upset polarity). The whole system consists of
three redundant cores, with majority-voted outputs. The internal scratchpad
RAMs of pBs from the three cores are majority-voted and scrubbed at
each processor reset, which is performed after a programmable number
of scrubbing cycles has been completed. BRAMs from the three cores
are continuously majority-voted and scrubbed via their second access
port. The configuration error detection runs in background with respect
to the logic implemented in the front-end FPGAs and it does not disrupt
operation. The $C^{3}$ is capable to correct any number of errors
per frame and it requires 3.3s to complete the parallel read back
of the 6 target FPGAs. Its resource occupation is just 1068 flip-flops,
2005 look-up-tables slices and 9 BRAMs, respectively 3\%, 6\% and
14\% of the overall available resources in the merger FPGA. The $C^{3}$
is designed for portability across most of the Xilinx families, from
the legacy Virtex-5 to the latest-generation Ultrascale+. 

\section{Test setup and results}

\begin{figure}
\begin{centering}
\includegraphics[width=1\columnwidth]{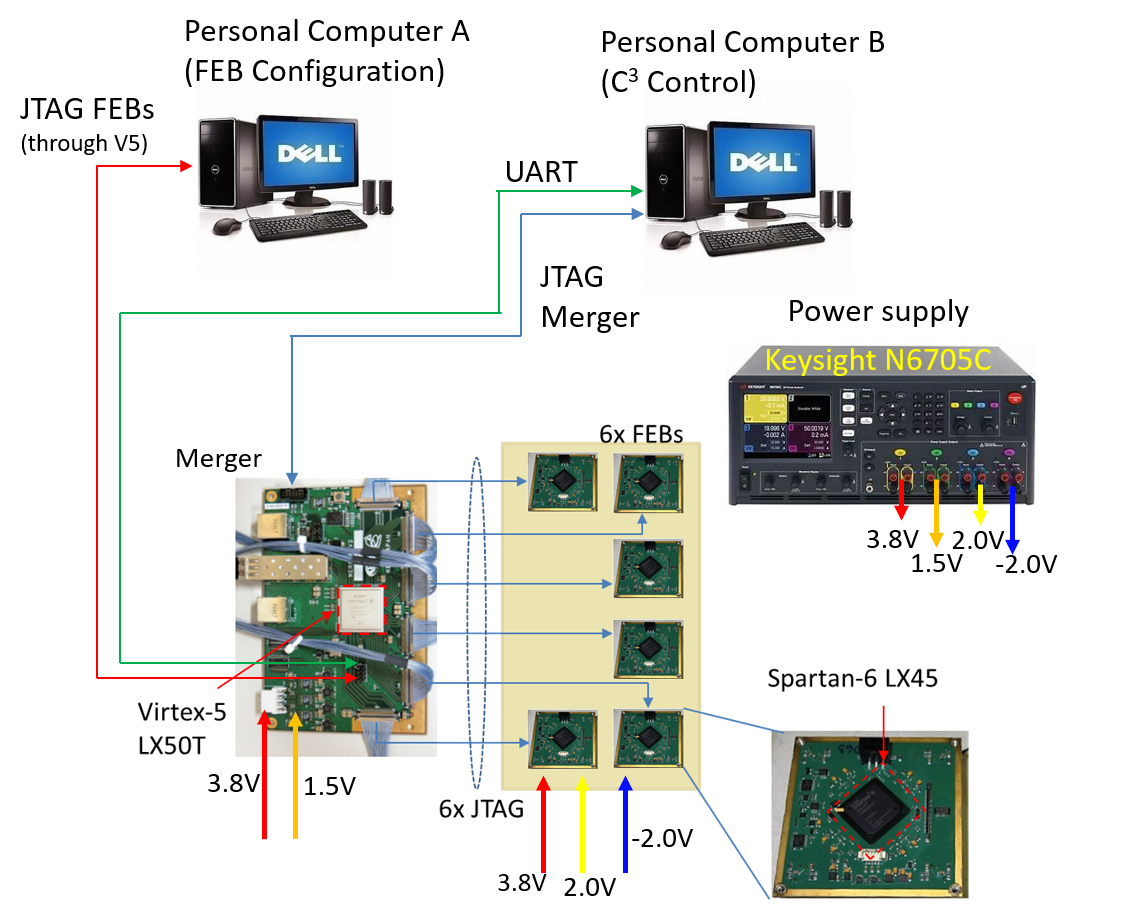}
\par\end{centering}
\caption{\label{fig:Fault_injection_test_setup}Fault-injection test setup
for the validation of the $C^{3}$. }
\end{figure}

We realized a test-setup (Fig. \ref{fig:Fault_injection_test_setup})
to verify the $C^{3}$ operation on the bench. The merger board is
initially configured as a pass-through to program the FEB FPGAs from
a dedicated personal computer (PC A). After configuration, another
personal computer (PC B) configures the merger with the $C^{3}$ bitstream.
At this point the PC B sends commands via UART to the $C^{3}$ to
inject SEUs and, after injection, it starts scrubbing. We injected
more than 4k upsets, uniformly distributed in devices, frames and
in the range from 1 to 4 upsets per frame. All the injected errors
have been detected and corrected. 
 We used a similar setup for a new neutron testing campaign of the
$C^{3}$ at the TRIGA reactor, in January 2020. Our results, show
that the $C^{3}$ effectively limits accumulation of upsets in configuration
memory of FEB FPGAs and it improves the mean time between failures
of the read out functionality by 30\% with respect to the SEM. During
the irradiation test we did not record any single-event latch-ups,
nor other hard failures, of the Merger and FEB FPGAs.

Moreover, the $C^{3}$ is operating in the ARICH since June 2020,
and SEU counts are logged via the EPICS-based slow control system.
We are using this data to monitor the SEU spatial distributions and
the SEU counts versus time for all the FEBs. We plan to study the
correlation of this data with the collider operating conditions.

\section*{Acknowledgment}

We wish to thank A. Boiano, A. Vanzanella, A. Pandalone, E. Masone
from SER (Electronics and Detectors Service) of Istituto Nazionale
di Fisica Nucleare Sezione di Napoli for their technical support to
this activity and the Jo\v{z}ef Stefan Institute TRIGA staff for
their technical support during the irradiation test. This work is
part of the ROAL project (grant no. RBSI14JOUV) funded by the Scientific
Independence of Young Researchers (SIR) 2014 program of the Italian
Ministry of Education, University and Research (MIUR).

\end{document}